\documentclass[10pt]{revtex4}
\usepackage{amssymb}
\usepackage{latexsym}
\usepackage{epsfig}
\begin{document}                             

\title{Interacting Ghost Dark Energy in Brans-Dicke Theory}

\author{Esmaeil Ebrahimi$^{1,2}$ \footnote{eebrahimi@uk.ac.ir} and Ahmad  Sheykhi$^{1,2,3}$ \footnote{
sheykhi@uk.ac.ir} }
\address{$^1$ Department of Physics, Shahid Bahonar University, PO Box 76175, Kerman, Iran\\
          $^2$ Research Institute for Astronomy and Astrophysics of Maragha (RIAAM), Maragha, Iran\\
          $3$ Physics Department and Biruni Observatory, Shiraz University, Shiraz 71454,
          Iran}

\begin{abstract}
We investigate the QCD ghost model of dark energy in the framework
of Brans-Dicke cosmology. First, we study the non-interacting ghost
dark energy in a flat Brans-Dicke theory. In this case we obtain the
EoS and the deceleration parameters and a differential equation
governing the evolution of ghost energy density. Interestingly
enough, we find  that the EoS parameter of the non-interacting ghost
dark energy can cross the phantom line ($w_D=-1$) provided the
parameters of the model are chosen suitably. Then, we generalize the
study to the interacting ghost dark energy in both flat and non-flat
Brans-Dicke framework and find out that the transition of $w_D$ to
phantom regime can be more easily achieved for than when resort to
the Einstein field equations is made.

\end{abstract}

 \maketitle

\section{Introduction}
Many observational evidences such as data from type Ia supernovae
\cite{Rie}, cosmic microwave background (CMB) \cite{cmb1,cmb2} and
SDSS \cite{sdss1,sdss2} have convinced people to accept that our
universe is currently experiencing a phase of accelerated expansion.
Based on the Einstein's theory of gravity we should introduce a new
type of energy with a negative pressure to provide such an epoch of
evolution. This unknown energy component which push the universe to
accelerate usually called dark energy (DE) in the literature. Of
course, the simplest candidate for DE is the cosmological constant
originally introduced by Einstein to explain the static behavior of
the universe at that time. However, cosmological constant  suffers
coincidence and fine tuning problems. Also a dynamic approach to
dark energy has been proposed avoiding such problems and in favor of
experimental evidences seeking a time varying equation of state
(EoS) parameter $w_D={p_D}/{\rho_D}$. Examples of such models are
quintessence \cite{wetter,ratra}, K-essence
\cite{chiba,armend1,armend2}, phantom \cite{cald}, quintom
\cite{feng}, tachyon \cite{tachyon}, holographic \cite{hol} and
agegraphic DE  \cite{age}.

In recent years there has been a new attention to the so called
``scalar-tensor gravity". Scalar-tensor models of gravity are those
introduce a scalar field modifying the Einstein's theory of gravity.
These models are mainly seen to retrieve at the low energy limit of
string theory. However, the history of these models back to many
years ago. The scalar tensor theories have started with the work by
P. Jordan in 1950 \cite{jordan}. A prototype of such models was
proposed by Brans and Dicke in 1961 \cite{BD}. Their aim for
presenting this model was to modify Einstein's theory in a way
admitting the so called "Mach's principle". To this end they
proposed a new scalar degree of freedom to incorporate the Mach's
principle into general relativity \cite{BD}. The Brans-Dicke (BD)
theory can pass the experimental tests from the solar system
\cite{bertotti} and provide an explanation for the acceleration of
the universe expansion \cite{mathiaz}.

Most DE models introduce new degrees of freedom in addition to those
exist in the standard model of cosmology. Introducing such degrees
of freedom need investigation about their nature and new
consequences in the universe. Hence, it seems so impressive and
economic if one can explain the DE puzzle using already presented
fluids and fields in the literature. Ghost dark energy (GDE) is an
example of these models which uses the so called Veneziano ghost
field to explain the recent acceleration of the universe
\cite{Urban,Ohta}. Originally the so called "Veneziano ghost field"
was presented as a solution to $U(1)$ problem in effective low
energy QCD \cite{witten,rosen,nath,kawar}. Taking into account the
ghost field leads some consequences in the vacuum energy density in
a dynamic spacetime or a spacetime with nontrivial topology, while
such a field seems to be un-physical in the Minkowski spacetime and
has no contribution to the vacuum energy density. In curved
spacetime the ghost field gives rise to a small vacuum energy
density proportional to $\Lambda^3_ {QCD} H$, where $H$ is the
Hubble parameter and $\Lambda^3_ {QCD}$ is QCD mass scale
\cite{Ohta}. It is shown that such a vacuum energy density is
capable to drive a phase of acceleration and can be considered as a
dynamical cosmological constant \cite{CaiGhost,sheykhigde}. It is
important to note that Freidmann equation for GDE model in Einstein
gravity is similar to the self-accelerating branch of the
Dvali-Gabadadze-Porrati (DGP) braneworld \cite{DGP,self1}, that is
both of them have the form $ H^2 \propto \alpha H +\rho_M,$
\cite{Ohta,def}. However we would like to emphasize here that in DGP
braneworld the term proportinal to $H$ appears due to the inclusion
of large extra dimension in the gravity theory and hence
modification of the Einstein field equations which yields an extra
degree of freedom in the theory, while GDE model is totally embedded
in standard model and general relativity, one needs not to introduce
any new parameter, new degree of freedom or to modify gravity. As a
result the origin of the term $\alpha H$ in GDE model completely
differs from previous models such as self-accelerating DGP
braneworld scenario \cite{def}.

Although, it is a general belief that the current curvature of the
universe is negligible and mostly the universe is considered with a
flat geometry, recent observations support the possibility of a
non-flat universe and detect a small deviation from $k=0$
\cite{nonflat1}. For example evidences from CMB and also supernova
measurements of the cubic correction to the luminosity distance
favor a positively curved universe \cite{nonflat2,nonflat3}. In
addition, some exact analysis of the WMAP data reveals the
possibility of a closed universe \cite{nonflat4}.

Various aspects of GDE have recently investigated. A generalization
of GDE model in a non-flat universe was discussed in \cite{feng}.
Tachyon and quintessence reconstruction of GDE model were studied in
\cite{tachgde,quintgde}. Since GDE model belongs to a dynamical
cosmological constant, it is more natural to study it in the
framework of BD theory than in Einstein gravity. In this paper, we
study a cosmological model of late acceleration based on the GDE
model in the framework of non-flat BD cosmology.

This paper is organized as follows. In the next section, we present
the GDE model in the flat BD theory. Interacting GDE model in a flat
BD theory  is  discussed in  section III. In section \ref{nonflat},
we generalize the study to the universe with spatial curvature. We
summarize our results in section \ref{sum}.

\section{Ghost dark energy in BD theory}\label{flat}

The action of BD theory, in the canonical form, can be written
\cite{Arik}
\begin{equation}
 S=\int{
d^{4}x\sqrt{g}\left(-\frac{1}{8\omega}\phi ^2
{R}+\frac{1}{2}g^{\mu \nu}\partial_{\mu}\phi \partial_{\nu}\phi
+L_M \right)},\label{act1}
\end{equation}
where ${R}$ is the scalar curvature and $\phi$ is the BD scalar
field. The non-minimal coupling term $\phi^2 R$  replaces with the
Einstein-Hilbert term ${R}/{G}$ in such a way that
$G^{-1}_{\mathrm{eff}}={2\pi \phi^2}/{\omega}$,  where
$G_{\mathrm{eff}}$ is the effective gravitational constant as long
as the dynamical scalar field $\phi$ varies slowly. In the Jordan
frame, the matter minimally couples to the metric and there is no
interaction between the scalar field $\phi$ and the matter fields.
The equations of motion for the metric $g_{\mu\nu}$ and the BD
scalar field $\phi$ are

\begin{eqnarray}
&&\phi \,G_{\mu\nu}=-8\pi T_{\mu\nu}^{M} - \frac{\omega}{\phi}
\left(\phi_{,\mu}\phi_{,\nu}-\frac{1}{2}g_{\mu\nu}\phi_{,\lambda}\phi^{,\lambda}\right)
-\phi_{;\mu;\nu}+g_{\mu\nu}\Box\phi, \label{eqn1}
\\ & &\Box\phi=\frac{8\pi}{2\omega+3}T_{\lambda}^{M
\,\lambda}, \label{eqn2}
\end{eqnarray}
where $T^M_{\mu\nu}$ is the matter energy-momentum tensor. Our aim
in this paper is to consider the GDE in the
Friedmann-Robertson-Walker (FRW) universe which is described by
the line element
\begin{eqnarray}
 ds^2=dt^2-a^2(t)\left(\frac{dr^2}{1-kr^2}+r^2d\Omega^2\right),\label{metric}
 \end{eqnarray}
where $a(t)$ is the scale factor, and $k$ is the curvature parameter
with $k = -1, 0, 1$ corresponding to open, flat, and closed
universes, respectively. A closed universe with a small positive
curvature ($\Omega_k\simeq0.01$) is compatible with observations
\cite{spe}. Using  metric (\ref{metric}), the field equations
(\ref{eqn1}) and (\ref{eqn2}) reduce to
\begin{eqnarray}
 &&\frac{3}{4\omega}\phi^2\left(H^2+\frac{k}{a^2}\right)-\frac{1}{2}\dot{\phi} ^2+\frac{3}{2\omega}H
 \dot{\phi}\phi=\rho_M+\rho_D,\label{FE1}\\
 &&\frac{-1}{4\omega}\phi^2\left(2\frac{{\ddot{a}}}{a}+H^2+\frac{k}{a^2}\right)-\frac{1}{\omega}H \dot{\phi}\phi -\frac{1}{2\omega}
 \ddot{\phi}\phi-\frac{1}{2}\left(1+\frac{1}{\omega}\right)\dot{\phi}^2=p_D,\label{FE2}\\
 &&\ddot{\phi}+3H
 \dot{\phi}-\frac{3}{2\omega}\left(\frac{{\ddot{a}}}{a}+H^2+\frac{k}{a^2}\right)\phi=0,
 \label{FE3}
\end{eqnarray}
where $H=\dot{a}/a$ is the Hubble parameter, $\rho_D$ and $p_D$ are,
respectively, the energy density and pressure of DE, and $\rho_M$ is
the pressureless dark matter (DM) density.

Consider the FRW universe filled with DE and pressureless matter
which evolves according to their conservation laws
\begin{eqnarray}
&&
\dot{\rho}_D+3H\rho_D(1+w_D)=0,\label{consq12}\\
&&\dot{\rho}_{M}+3H\rho_{M}=0, \label{consm12}
\end{eqnarray}
where $w_D$ is the EoS parameter of DE. In this section we want to
consider the GDE in a spatially flat spacetime in the BD framework.
The ghost energy density is proportional to the Hubble parameter
\cite{Ohta}
\begin{equation}\label{GDE}
\rho_D=\alpha H.
\end{equation}
Here $\alpha$ is a constant of order $\Lambda_{\rm QCD}^3$ where
$\Lambda_{\rm QCD}\sim 100 MeV$ is QCD mass scale. With
 $H \sim 10^{-33}eV$,
$\Lambda^3_{\rm QCD}H$ gives the right order of magnitude $\sim
(3\times10^{-3}eV)^4$ for the observed DE density \cite{Ohta}.

To determine the evolution of the universe filled by the
pressureless matter and GDE, through the equations
(\ref{FE1}-\ref{FE3}), we still have another degree of freedom in
analyzing the set of equations. Based on the previous experiences in
the BD theory it is known that usually the BD scalar field $\phi$
has a power law relation as
\begin{equation}\label{phipl}
\phi=\phi_0a(t)^{\varepsilon}.
\end{equation}
A case of particular interest is that when $\varepsilon$ is small
whereas $\omega$ is high so that the product $\varepsilon\omega$
results of order unity \cite{banpavon}. This is interesting because
local astronomical experiments set a very high lower bound on
$\omega$ \cite{Will}; in particular, the Cassini experiment implies
that $\omega
> 10^4$ \cite{bertotti,acqui}. Taking the derivative with respect to time of
relation (\ref{phipl}) we obtain
\begin{equation}\label{phidot}
 \frac{\dot{\phi}}{\phi}=\varepsilon \frac{\dot{a}}{a}=\varepsilon H.
\end{equation}
Using Eqs. (\ref{phipl}) and (\ref{phidot}), the first Friedmann
equation (\ref{FE1}) becomes
\begin{equation}\label{frid1}
  H^2(1-\frac{2\omega}{3}\varepsilon^2+2\varepsilon)+\frac{k}{a^2}=\frac{4\omega}{3\phi^2}(\rho_D+\rho_M).
\end{equation}
We introduce the fractional energy densities corresponding to each
energy component as usual
\begin{eqnarray}
\Omega_M&=&\frac{\rho_M}{\rho_{\mathrm{cr}}}=\frac{4\omega\rho_M}{3\phi^2
H^2}, \label{Omegam} \\
\Omega_k&=&\frac{\rho_k}{\rho_{\mathrm{cr}}}=\frac{k}{H^2 a^2},\label{Omegak} \\
\Omega_D&=&\frac{\rho_D}{\rho_{\mathrm{cr}}}=\frac{4\omega\rho_D}{3\phi^2
H^2}, \label{OmegaD}
\end{eqnarray}
where we have defined
\begin{eqnarray}\label{rhocr}
\rho_{\mathrm{cr}}=\frac{3\phi^2 H^2}{4\omega}.
\end{eqnarray}
Using (\ref{GDE}) we can rewrite Eq. (\ref{OmegaD}) as
\begin{equation}\label{OmegaD2}
 \Omega_D=\frac{4\omega\alpha}{3\phi^2H}.
\end{equation}
Based on these definitions, Eq. (\ref{frid1}) can be rewritten as
\begin{equation}\label{frid1omega}
   \gamma=\Omega_D+\Omega_M-\Omega_k,
\end{equation}
where
\begin{equation}
\gamma=1-\frac{2\omega}{3}\varepsilon^2+2\varepsilon.
\end{equation}
Taking the time derivative of relation (\ref{GDE}), and using the
Friedmann equation (\ref{frid1}) as well as the continuity equations
(\ref{consq12}) and (\ref{consm12}), we find
\begin{equation}\label{eq1}
  \frac{\dot{\rho}_D}{\rho_D}=\frac{\dot{H}}{H}=-H\left[\varepsilon+\frac{3}{2}+\frac{3}{2}\frac{\Omega_D
    w_D}{\gamma}\right].
\end{equation}
Substituting this relation in continuity equation (\ref{consq12}) we
obtain the EoS parameter of GDE, namely
\begin{equation}\label{wflat}
   w_D=\frac{\gamma}{2\gamma-\Omega_D}\left[\frac{2\varepsilon}{3}-1\right].
\end{equation}
It is worthy to note that in the limiting case $\varepsilon=0$
($\omega\rightarrow\infty$) we have $\gamma=1$ and hence the BD
scalar field becomes trivial; as a result Eq. (\ref{wflat}) reduces
to its respective expression in flat standard cosmology
\cite{sheykhigde}
\begin{eqnarray}
w_D=-\frac{1}{2-\Omega_D}.\label{wDstand}
\end{eqnarray}
\begin{figure}\epsfysize=5cm
{ \epsfbox{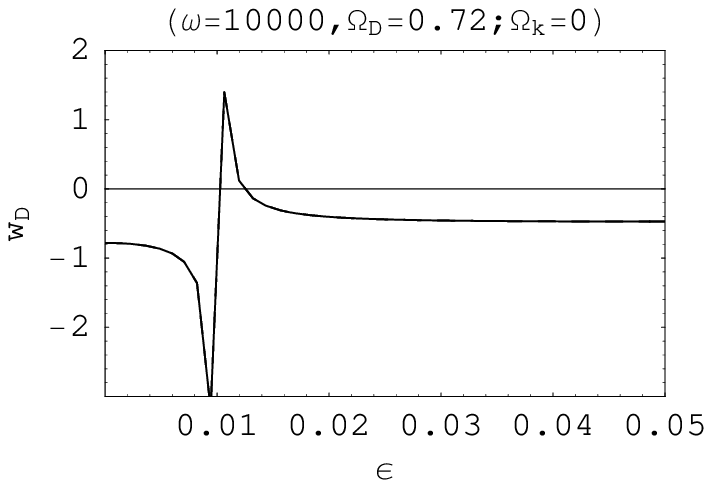}}\epsfysize=5cm
{\epsfbox{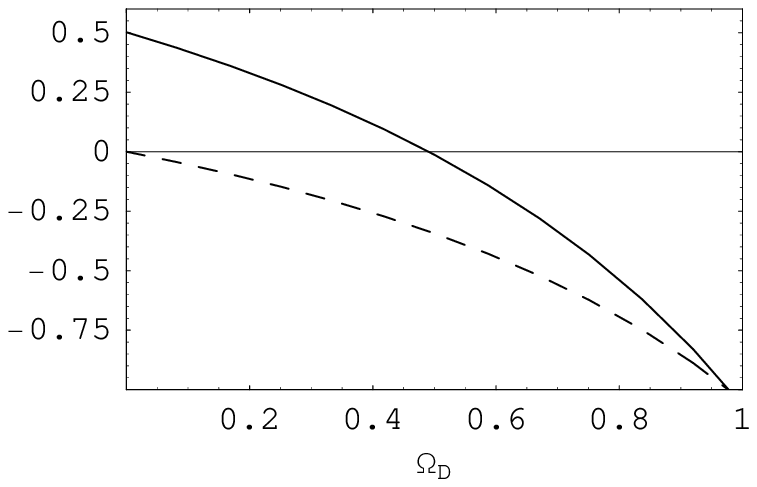}}\caption{The first figure indicates $w_D$
against $\varepsilon$ in the flat non-interacting GDE. In the second
figure the deceleration parameter (solid) and $w_{\rm eff}$ (dashed)
are plotted versus $\Omega_D$ for $\varepsilon=0.002$.} \label{i1}
\end{figure}
The solar-system experiments give the result for the value of
$\omega$ is $\omega > 40000$ \cite{bertotti}. However, when probing
the larger scales, the limit obtained will be weaker than this
result. In Ref. \cite{acqui}, the authors found that $\omega$ is
smaller than 40000 on the cosmological scales. Also, Wu and Chen
\cite{wu} obtained the observational constraints on BD model in a
flat universe with cosmological constant and cold DM using the
latest WMAP and SDSS data. They found that within $2\sigma$ range,
the value of $\omega$ satisfies $\omega<-120.0$ or $\omega>97.8$
\cite{wu}. They also obtained the constraint on the rate of change
of $G$ at present
\begin{equation}\label{exp1}
    -1.75 \times 10^{-12} yr^{-1} <
\frac{\dot{G}}{G} < 1.05 \times 10^{-12} yr^{-1}
\end{equation}
at $2\sigma$ confidence level. So in our case with  assumption
(\ref{phipl}) we get
\begin{equation}\label{exp2}
\frac{\dot{G}}{G}=\frac{\dot{\phi}}{\phi}=\varepsilon H< 1.05
\times 10^{-12} yr^{-1}
\end{equation}
 This relation can be used to put an upper bound on
 $\varepsilon$. Assuming the present value of the Hubble
 parameter to be $H_{0}\simeq 0.7$ we obtain
 \begin{equation}\label{exp3}
    \varepsilon<0.01.
\end{equation}

The GDE model in BD framework has an interesting feature compared to
the GDE model in Einstein's gravity. It was shown that in standard
cosmology based on Einstein's theory, the EoS parameter of the
noninteracting GDE cannot cross the phantom line $w_D=-1$ and at the
late time where $\Omega_D\rightarrow1$ approaches
$-1$\cite{sheykhigde}. However, in the BD framework, requiring
condition $w_D<-1$ leads to
$\gamma(\frac{2\varepsilon}{3}+1)<\Omega_D$. Choosing
$\Omega_D=0.72$ for the present time, this inequality valid provided
we take $\varepsilon=0.002$ which is consistent with observations.
This indicates that one can generate a phantom-like EoS for the
noninteracting GDE in the BD framework.

In addition to the EoS parameter of the GDE we can also study the
effective EoS parameter, $w_{\rm eff}$, which is defined as
\begin{equation}\label{wef}
    w_{\rm eff}=\frac{P_t}{\rho_t}=\frac{P_D}{\rho_D+\rho_M},
\end{equation}
where $\rho_t$ and $P_t$ are, respectively, the total energy density
and total pressure of the universe. As usual we assumed the dark
matter is in the form pressureless fluid ($P_M$=0). Using relation
(\ref{frid1omega}) for the flat case one can find

\begin{equation}\label{wef2}
    w_{\rm eff}=\frac{\Omega_D}{\gamma}w_D=\frac{\Omega_D}{2\gamma-\Omega_D}\left[\frac{2\varepsilon}{3}-1\right].
\end{equation}

It is also interesting to study the behavior of the deceleration
parameter defined as
\begin{eqnarray}\label{q}
q=-\frac{\ddot{a}}{aH^2}=-1-\frac{\dot{H}}{H^2}.
\end{eqnarray}

Inserting from (\ref{eq1}) into  (\ref{q}) yields
\begin{equation}\label{qpf}
    q=\frac{1}{2}+\varepsilon+\frac{3}{2}\frac{\Omega_D
    w_D}{\gamma}
\end{equation}
Substituting Eq. (\ref{wflat}) in the above relation one can easily
reach
\begin{equation}\label{qf}
    q=\frac{1}{2}+\varepsilon+\frac{3\Omega_D}{2(2\gamma-\Omega_D)}
    \left(\frac{2\varepsilon}{3}-1\right).
\end{equation}
Let us study some special cases of interest for the deceleration
parameter $q$. If we take $\Omega_D=0.72$ for the present time and
choosing $\varepsilon=0.002$ and $\omega=10^4$ we obtain $q=-0.36$,
which is consistent with the present value of the deceleration
parameter obtained in \cite{Daly}. This choice of parameters lead
$w_D=-0.79$ and $w_{\rm eff}=-0.58$. The evolution of the
deceleration parameter $q$ and $w_{\rm eff}$ are plotted in the
second part of Fig(\ref{i1}). A close look to this figure reveals
that the universe enters acceleration phase when $\Omega_D=0.49$.
The effective EoS parameter at $\Omega_D=0.49$ becomes  $w_{\rm
eff}=-0.33$ while $w_D=-0.66$. We can also study the behavior of the
deceleration parameter in the early stage of the universe where
$\Omega_D\ll 1$. In this epoch $q=\frac{1}{2}+\varepsilon$ which
indicates that the universe was experiencing a phase of deceleration
at the early stage of its evolution due to the domination of the DM
component.

Once again we can see that in the limiting case $\varepsilon=0$
($\gamma=1)$ the above relation restores the deceleration parameter
of GDE in Einstein's gravity \cite{sheykhigde}.
\begin{equation}\label{qfahmad}
    q=\frac{1}{2}-\frac{3}{2}\frac{\Omega_D}{(2-\Omega_D)}.
\end{equation}
Finally, we obtain a differential equation governing the evolution
of GDE from early deceleration to late time acceleration. To do this
we take time derivative of the relation (\ref{OmegaD2}) and use Eq.
(\ref{q}). The result is
\begin{equation}\label{omegaDdot}
    \dot{\Omega}_D=\Omega_D H(1+q-2\varepsilon).
\end{equation}
Inserting $q$ from (\ref{qf}) and using relation
$\Omega_D^{\prime}=H\frac{d\Omega_D}{d\ln{a}}$, we obtain
\begin{equation}\label{OmegaDev}
\Omega^{\prime}_D=\Omega_D\left[3\left(\frac{1-\Omega_D}{2\gamma-\Omega_D}
\right)-2\varepsilon\right],
\end{equation}
where prime denotes the derivation with respect to $x=\ln a$. In the
limiting case $\varepsilon=0$ ($\gamma=1)$ one recovers the result
obtained in \cite{sheykhigde}.
\section{Interacting Ghost dark energy in flat BD
theory}\label{flat}

Most of the models study the dark side of the universe consider
the evolution of DE and DM separately. This means that $\rho_{DM}$
and $\rho_{DE}$ are separately conserved. However, recently there
has been a lot of interest in interacting models of DE, since
observations detects a signal of interaction between DE and DM.
For instance, observational evidences provided by the galaxy
cluster Abell A586 supports the interaction between DE and DM
\cite{interact1}. Beside there is no any reason against the
interacting behavior of DE and DM and many authors has explained
Lagrangians leaded to an interacting approach \cite{samisuji}. Any
conservation in the physics can be explained by a symmetry in the
Lagrangian level and there does not exist such a symmetry through
the known Lagragians explaining the dark side of the universe.
These points convince us to discuss a version of GDE in which
there exist an interaction between DE and DM. To this end we can
write the conservation equation for different components as
\begin{eqnarray}
&&
\dot{\rho}_D+3H\rho_D(1+w_D)=-Q,\label{consq2}\\
&&\dot{\rho}_{M}+3H\rho_{M}=Q, \label{consm2}
\end{eqnarray}
where $Q$ denotes the interaction term and we take it as $Q =3b^2
H(\rho_{M}+\rho_{D})$ with $b^2$ is a coupling constant. Such a
choice for interacting term implies the the DE and DM component do
not conserve separately while the total density is still conserved
through
\begin{equation}\label{totcons}
\dot{\rho}+3H(\rho+P)=0,
\end{equation}
where $\rho=\rho_D+\rho_M$ and $P=P_D$.
\begin{figure}\epsfysize=5cm
{ \epsfbox{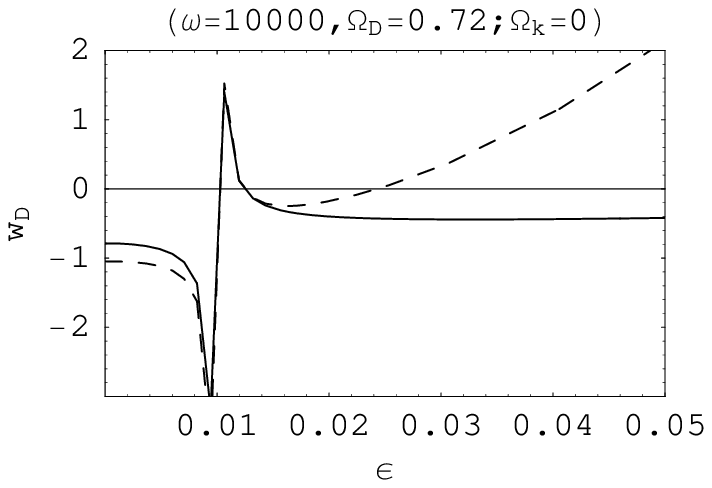}}\epsfysize=5cm {
\epsfbox{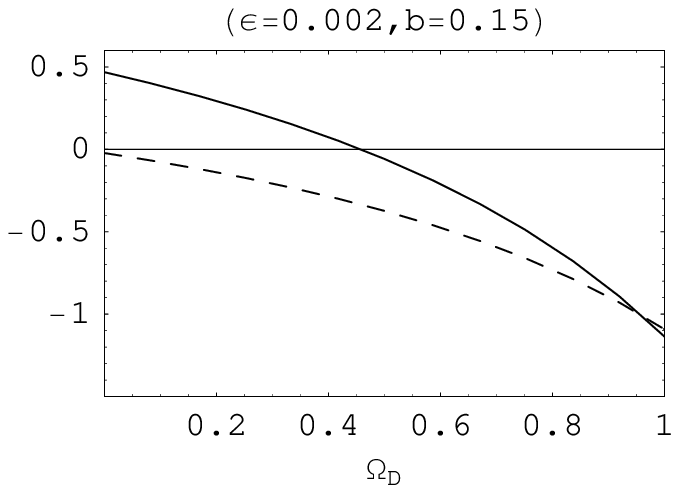}}\caption{In the first figure $w_D$ is plotted
against $\varepsilon$. Solid line corresponds to $b=0.05$ and dashed
one for $b=0.2$. In the second figure the deceleration parameter
(solid) and the $w_{\rm eff}$ (dashed) are plotted versus
$\Omega_D$. These figures are plotted in the flat, interacting GDE
case. } \label{i2}
\end{figure}
In this section we would like to discuss the interacting GDE in a
flat universe. To this goal, once again we take time derivative
from (\ref{GDE}) to get (\ref{eq1}). Next, we obtain the L.H.S. of
Eq. (\ref{eq1}) from the conservation equation and the R.H.S. from
(\ref{frid1}) and also taking into account $Q =3b^2
H(\rho_{M}+\rho_{D})$ and (\ref{frid1omega}) we obtain the EoS
parameter
\begin{equation}\label{wflat2}
 w_D=\frac{\gamma}{2\gamma-\Omega_D}\left[\frac{2\varepsilon}{3}-1-\frac{2\gamma b^2}{\Omega_D}\right].
\end{equation}
In the limiting case $\varepsilon=0$ ($\omega\rightarrow\infty$) or
$\gamma=1$, the BD theory reduces to the Einstein gravity and hence
Eq. (\ref{wflat2}) recovers its respective expression in standard
cosmology \cite{sheykhigde}
\begin{eqnarray}
w_D=-\frac{1}{2-\Omega_D}(1+\frac{2b^2}{\Omega_D}).\label{wDstand2}
\end{eqnarray}

The effective EoS parameter can be written as
\begin{equation}\label{wefflatin}
w_{\rm
eff}=\frac{\Omega_D}{\gamma}w_D=\frac{\Omega_D}{2\gamma-\Omega_D}\left[\frac{2\varepsilon}{3}-1-\frac{2\gamma
b^2}{\Omega_D}\right].
\end{equation}

In this case, using the method of the previous section, the
deceleration parameter is obtained as
\begin{equation}\label{qf2}
    q=\frac{1}{2}+\varepsilon+\frac{3\Omega_D}{2(2\gamma-\Omega_D)}
    \left[\frac{2\varepsilon}{3}-\left(1+\frac{2\gamma
    b^2}{\Omega_D}\right)\right].
\end{equation}
In order to obtain an insight about various features of the
interacting case, we try a suitable choice of parameters. Taking
$\Omega_D=0.72, b=0.15$ and $\varepsilon=0.002$, we obtain $q=-0.41$
for the present value of the deceleration parameter which is in good
agreement with recent observational results \cite{Daly}. Also this
choice of the parameters result in $w_D=-0.84$ and $w_{\rm
eff}=-0.62$ for present time of the universe and one can see the
consistency of the latter with current observations. Transition from
deceleration to acceleration take places at $\Omega_D=0.46$ and at
this time $w_{\rm eff}=-0.33$ while $w_D=-0.71$. One may take a look
on Fig. \ref{i2} for a better insight into the model. The first part
of this figure indicates that one can generate a phantom-like
behavior provided $\varepsilon<0.01$. Further, we can have a close
look at the behavior of $w_D$ at the late time where
$\Omega_D\rightarrow 1$. In this limit and using the same values of
$\varepsilon$ and $b$ we find $w_D=-1.06$, indicating that the
interacting GDE in the BD framework can cross the phantom line in
the future. In this limit with a same set of parameters $w_{\rm
eff}=-1.09$ which leads a super acceleration and may result in a big
rip as fate of the universe.

The equation of motion of GDE is obtained as
\begin{equation}\label{OmegaDev2}
\Omega^{\prime}_D=\Omega_D\left[\frac{3}{2}\left(1-\frac{\Omega_D}{2\gamma-\Omega_D}
    \left(1+\frac{2\gamma
    b^2}{\Omega_D}\right)\right)-2\varepsilon\right].
\end{equation}

\section{Interacting Ghost dark energy in non-flat universe}\label{nonflat}
Historically, one of the problems leaded to pioneering theory of
inflation was the so-called ``flatness problem". flatness problem
was explained by assuming a very huge rate of expansion in the first
stages of the cosmic evolution namely ``inflation". After the
inflation it is generally believed that the universe is spatially
flat, that is the inflation practically washes out the effect of
curvature in the early stages of cosmic evolution. However, in the
past decade several observational evidences have been observed in
contrast to the flatness assumption of the universe. In the context
of inflation it is argued that the flatness is not a necessary
consequence of inflation if the number of e-foldings is not very
large \cite{huang}. Besides, the parameter $\Omega_k$ represents the
contribution to the total energy density from the spatial curvature
and it is constrained as $-0.0175 <\Omega_k< 0.0085$ with $95\%$
confidence level by current observations \cite{water}. In addition,
there are several observational evidences which supports a closed
geometry for our universe in
\cite{nonflat1,nonflat2,nonflat3,nonflat4}. Based on these
implications it seems interesting to investigate the different DE
models while the curvature term is present. So, in this section our
main aim is to discuss a non-flat FRW background in the presence of
the interacting GDE in BD theory. Taking the curvature into account,
the Friedmann equation is rewritten as
\begin{eqnarray}\label{Friedm}
\gamma H^2+\frac{k}{a^2}=\frac{4\omega}{3\phi^2} \left(
\rho_m+\rho_D \right).
\end{eqnarray}

\begin{figure}\epsfysize=5cm
{ \epsfbox{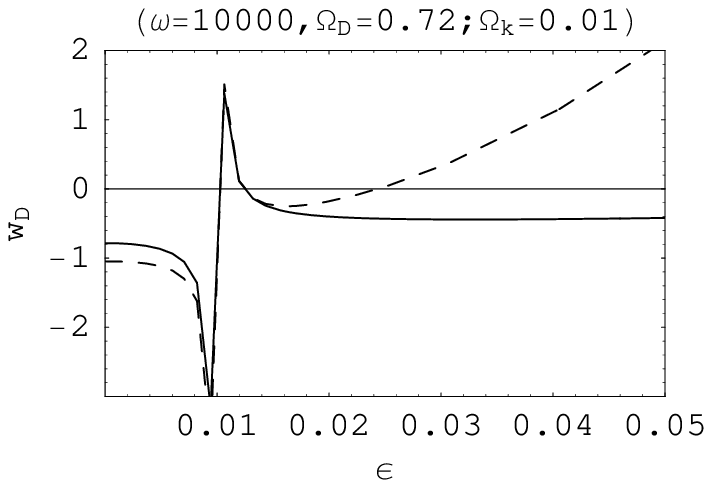}}\epsfysize=5cm {
\epsfbox{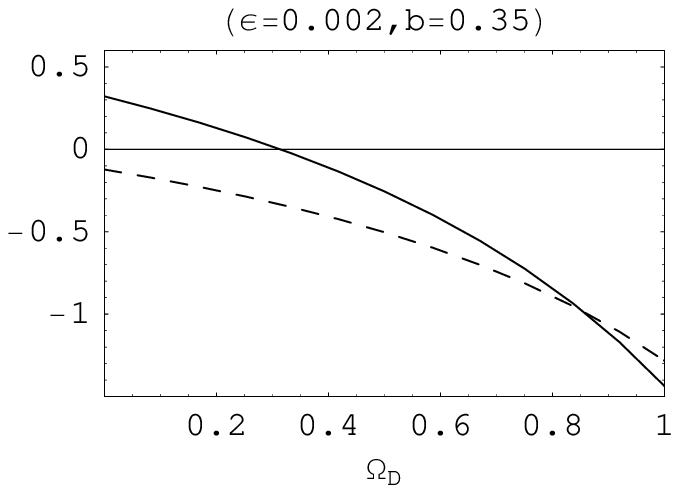}}\caption{In this figure $w_D$ is plotted
against $\varepsilon$. Solid line corresponds to $b=0.05$ and dashed
one for $b=0.2$. The second figure shows the evolution of
deceleration parameter (solid) and also $w_{\rm eff}$ (dashed)
versus $\Omega_D$. These figures are plotted in the presence of the
curvature as well as the interaction between DE and DM for GDE in BD
framework.} \label{i3}
\end{figure}

Following the method of the previous section we obtain different
parameters in the presence of the curvature term. Taking the time
derivative of the Friedmann equation (\ref{Friedm}) and using
(\ref{frid1omega}), we find
\begin{equation}\label{doth2}
\frac{\dot{H}}{H^2}=\frac{\Omega_k}{\gamma}-(1+\frac{\Omega_k}{\gamma})
\left[\varepsilon+\frac{3}{2}+\frac{3}{2}\frac{\Omega_Dw_D}{\gamma+\Omega_k}\right].
\end{equation}
Combining the above equation with Eq. (\ref{consq2}), after using
Eq. (\ref{eq1}), we obtain the EoS parameter as
\begin{equation}\label{wDn2}
w_D=-\frac{\gamma}{2\gamma-\Omega_D}\left(1-\frac{\Omega_k}{3\gamma}-\frac{2\varepsilon}{3}(1+\frac{\Omega_k}{\gamma})
+\frac{2b^2}{\Omega_D} (\gamma+\Omega_k)\right).
\end{equation}
One can easily check that the above equation reduces to the EoS
parameter of non-flat interacting GDE in the Einstein's framework
provided $\varepsilon=0$ ($\gamma=1$).

Using the same method as the previous sections, $w_{\rm eff}$ in a
non-flat background reads
\begin{equation}\label{wefnfint}
w_{\rm eff}=\frac{\Omega_D}{\gamma+\Omega_k}w_D.
\end{equation}
This relation can be obtained using Eq.(\ref{wef}) as well as
(\ref{frid1omega}) in the non-flat case. Replacing (\ref{wDn2}) in
the above relation, we get
\begin{equation}\label{wefnfint}
w_{eff}=-\frac{\gamma\Omega_D}{(2\gamma-\Omega_D)(\gamma+\Omega_k)}\left(1-\frac{\Omega_k}{3\gamma}-\frac{2\varepsilon}{3}(1+\frac{\Omega_k}{\gamma})
+\frac{2b^2}{\Omega_D} (\gamma+\Omega_k)\right).
\end{equation}
The deceleration parameter can be calculated by substituting Eqs.
(\ref{doth2}) and (\ref{wDn2}) into (\ref{q}). We find
\begin{equation}\label{q2}
q=(1+\frac{\Omega_k}{\gamma})\left[\frac{1}{2}-\frac{2\gamma\varepsilon}{\Omega_D-2\gamma}\right]
+\frac{3\Omega_D}{2(2\gamma-\Omega_D)}\left[1-\frac{\Omega_k}{3\gamma}+\frac{2b^2}{\Omega_D}(\gamma+\Omega_k)\right].
\end{equation}
Also the equation of motion of GDE can be obtained by replacing Eq.
(\ref{q2}) in (\ref{omegaDdot}). The result is
\begin{equation}\label{Omegaprime2n}
\frac{d\Omega_D}{d\ln
a}=\Omega_D\left[1+(1+\frac{\Omega_k}{\gamma})\left[\frac{1}{2}-\frac{2\gamma\varepsilon}{\Omega_D-2\gamma}\right]
+\frac{3\Omega_D}{2(2\gamma-\Omega_D)}\left[1-\frac{\Omega_k}{3\gamma}+\frac{2b^2}{\Omega_D}(\gamma+\Omega_k)\right]-2\varepsilon\right].
\end{equation}
It is worth to mentioning that all relations we obtained in this
section restor their respective expressions in the previous
section when we set $\Omega_k=0$ as we expect. Furthermore, one
can see that the evolution equation of the GDE and the
deceleration parameter reduce to the expected formulas in the
Einstein's theory \cite{sheykhigde}. A close look to
Fig.(\ref{i3}) reveals some interesting features of the model.
First of all, we can see that the EoS parameter, $w_D$ in all of
the figures just can result in an acceleration phase if
$\varepsilon<0.01$ which this point is completely consistent with
the observational evidences \cite{wu}. Additionally one can note
that in all  cases including flat,  non flat, interacting and
noninteracting versions of the GDE model, the EoS parameter of GDE
can achieve the phantom cross with suitable choices of
$\varepsilon$ and the interaction coupling parameter $b$. As an
example taking $\Omega_k=0.01$ favored by observation,
$\Omega_D=0.72$ for the present time, $\varepsilon=0.002$ and
$b>0.32$ lead to $w_D<-1.01$. It is worth noting that the
existence of the spatial curvature in this section does not lead
to a considerable difference between the acceleration rate or the
deceleration rate of the universe with respect to the flat case.
However, it should be emphasize that the existence of a non-zero
curvature in the universe has very important consequences on
global topology of the universe and much subtle issues.
\section{Discussion}\label{sum}
Recently, the vacuum energy of the Veneziano ghost field in a
time-dependent background was proposed as a kind of DE candidate to
explain the acceleration of the cosmic expansion. In this model, the
energy density of the dark energy is proportional to the Hubble
parameter $H$. In this paper we investigated GDE model in the BD
framework. In this framework we studied the GDE in three distinct
cases. At first, we generalized the GDE to BD theory in the absence
of interaction between DM and GDE in a flat FRW background. In this
case the EoS parameter was obtained and it was found that for
$\Omega_D=0.72$, $\omega=10000$ and $\varepsilon=0.002$ the universe
enters a phase of accelerated expansion at the late time. Another
interesting result from this part is that one can generate a
phantom-like EoS from a noninteracting GDE in a flat background,
which is in contrast to the GDE in standard cosmology. For example,
taking $\Omega_D=0.72$, $\omega=10000$ and $\varepsilon=0.007$ leads
$w_D=-1.04$. In the next two sections we presented the interacting
GDE in both flat and non-flat BD cosmology. In both of these
sections we found that the universe can enter a phase of
acceleration as well as it can cross the phantom line with suitable
choice of constants. For example for $\Omega_k=0.01$ favored by
observation, $\Omega_D=0.72$ for the present time,
$\varepsilon=0.002$ and $b>0.35$ lead to $w_D<-1.05$. For all cases
we plotted $w_D$ versus $\varepsilon$ and found that crossing from
the phantom line is possible provided we take $\varepsilon<0.01$ and
in the region with $\varepsilon>0.01$ we sometimes even cannot have
a phase of acceleration. The important point is that such a
constraint exist from the observational point of view \cite{wu} and
we obtained such a condition theoretically. Also we investigated the
deceleration parameter in different cases. For flat universe in the
absence of interaction between DE and DM we found that with
$\Omega_D=0.72$ for present time and choosing $\varepsilon=0.002$
and $\omega=10000$, we reach $q=-0.36$ which shows a consistent
value in comparison with observation. Also this parameter have a
consistent behavior in all cases of interest.

\acknowledgments{This work has been supported by Research Institute
for Astronomy and Astrophysics of Maragha, Iran.}

\end{document}